\documentstyle[aps,epsf]{revtex}
\twocolumn
\begin{document}
\title{Single-Electron Traps: A Quantitative Comparison of Theory and Experiment}
\author{K.A. Matsuoka,  K.K. Likharev, P. Dresselhaus, L. Ji, S. Han, and J. Lukens}
\address{Department of Physics, State University of New York,\\
Stony Brook, NY 11794-3800}
\maketitle

\begin{abstract}
We have carried out a coordinated experimental and theoretical study of
single-electron traps based on submicron metallic (aluminum) islands and
Al/AlO$_x$/Al tunnel junctions.  The results of geometrical modeling using
a modified version of MIT's FastCap were used as input data for the
general-purpose single-electron circuit simulator {\sc moses}.  The
analysis indicates reasonable quantitative agreement between theory and
experiment for those trap characteristics which are not affected by random
offset charges.  The observed differences (ranging from a few to fifty
percent) can be readily explained by the uncertainty in the exact geometry
of the experimental nanostructures.
\end{abstract}


\section{Introduction}

Recent advances in the physics of single-electron charging of macroscopic
conductors (for general reviews see, e.g., Refs. \onlinecite{mes,sct}) have
led to proposals for several new analog and digital electronic devices.
Such devices are considered, in particular, to be the most likely
candidates to replace silicon transistors in future ultra-dense electronic
circuits -- see, e.g., Refs.~\onlinecite{pasct,sasha}.

Single-electronics is presently one of the most active areas of solid state
physics and electronics, with hundreds of experimental and theoretical
works being published annually.  We are not aware, however, of any previous
attempts to quantitatively compare experimental data for a particular
device with results of theoretical analysis including geometrical
modeling\cite{knoll}.  Such a comparison was the main objective of
this work.  To that end, we selected one of the simplest devices, the
single-electron trap\cite{pasct,fulton}.

\begin{figure}
\begin{center}
\leavevmode
\epsfxsize=6cm
\epsffile{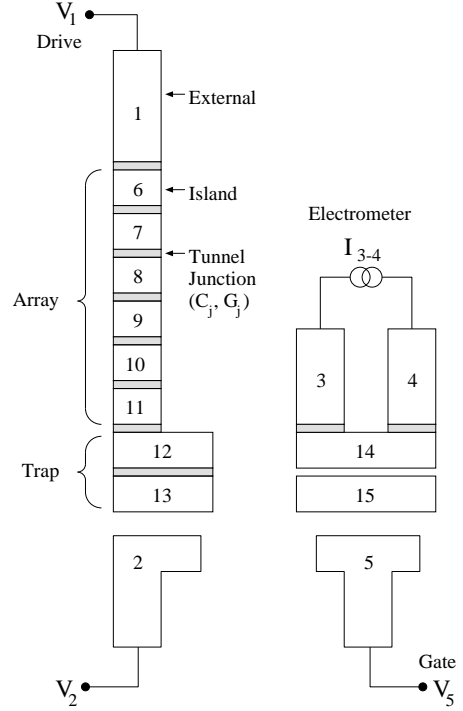}
\end{center}
\caption{Schematic view of the 8-junction single-electron trap/electrometer
circuit.  Islands 12 and 13 are strongly coupled and together provide the
energy well for an extra electron.  Islands 6-11 form the array separating
the well from the drive external 1.}
\label{schematic}
\end{figure}

Figure~\ref{schematic} shows the schematic layout of the circuit we discuss
in this paper, which consists of a trap coupled to a single-electron
electrometer. We will distinguish two types of conductors (``nodes'') in
the circuit: {\em externals}, wires which extend to the edges of the chip
and connect to the external measuring devices; and {\em islands}, small
metallic segments that are connected each other and to the externals by
tunnel junctions.

The trap consists of a larger island, providing the potential well for the
extra electron, separated from a voltage-biased ``drive'' external by an
$N$-island array.  The islands of the array are linked by $(N-1)$ tunnel
junctions with low capacitance $C_j$ and tunnel conductance $G_j$:

\begin{eqnarray}
C_j&\ll&\frac{e^2}{k_B T}\,, \label{lowc} \\
G_j&\ll&\frac{e^2}{h}\,.  \label{lowg}
\end{eqnarray}
\noindent

\begin{figure}
\begin{center}
\leavevmode
\epsfxsize=8cm
\epsffile{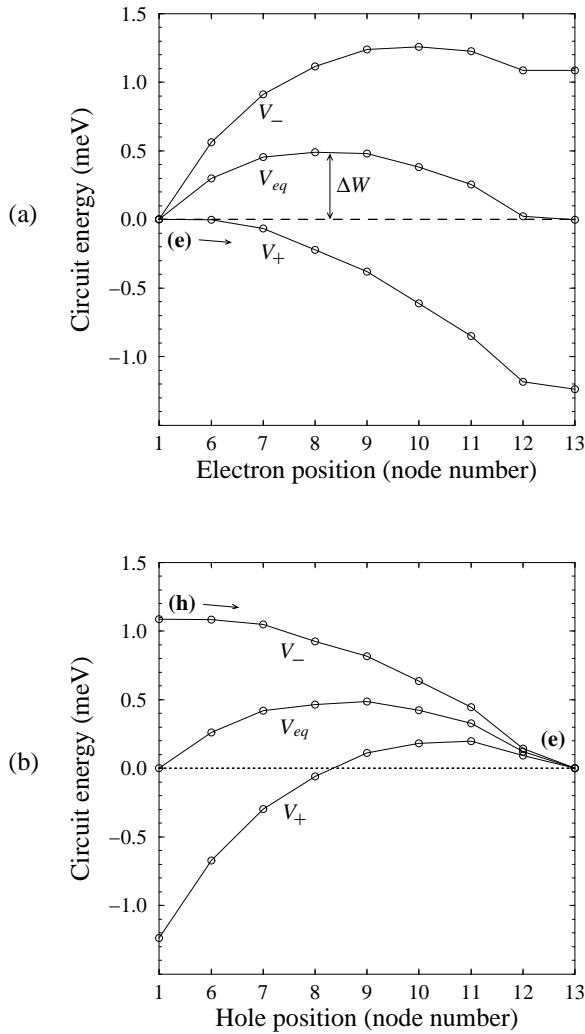}
\end{center}
\caption{Energy of a typical circuit calculated for three values of $V_1$
($V_+$, $V_{eq}$, and $V_-$), with $V_2$ = 8 mV: (a) as a function of the
position of one electron in the array; (b) as a function of the position of
one hole in the array, with one electron in the trap (node 13).}
\label{profile}
\end{figure}

Under condition (\ref{lowg}), each electron is localized inside a single
island at any given time. As Fig.~\ref{profile}a shows, the array creates
an electrostatic energy barrier $\Delta W\sim~e^2/C_j$ between the
drive electrode and the trap island.  To inject an additional electron
into the trap, a bias voltage $V=V_1-V_2$ is applied to the device.  At a
certain value $V=V_{+}$ the energy barrier is suppressed: an electron
tunnels from the drive external through the array and into the trap island.
To extract the electron from the trap island, a voltage $V=V_{-}$ is
applied, causing a hole to tunnel from the drive external to the trap
island, annihilating the trapped electron.

Electrons can also overcome the energy barrier by thermal activation and by
macroscopic quantum tunneling of charge (``cotunneling'').  At sufficiently
low temperatures (\ref{lowc}), the rate of thermally activated hopping over
the barrier is roughly~\cite{mes,pasct}

\begin{equation}
\label{gammat}\Gamma_T \sim \frac{G_j}{C_j}
\exp\left(\frac{-\Delta W}{k_B T}\right)\,,
\end{equation}

\noindent
while the rate of spontaneous cotunneling through the barrier scales as
\cite{mqt}

\begin{equation}
\label{gammaq}\Gamma _Q\sim \frac{G_j}{C_j}\left(\frac{G_jh}{4\pi ^2e^2}\right)
^{N-2}\,. 
\end{equation}
\noindent
If conditions (\ref{lowc}) and (\ref{lowg}) are satisfied, and the number
$N$ of junctions in the array is large enough, the rates of thermal
activation and cotunneling may be very low.  Thus, the lifetime $\tau_L =
(\Gamma_T+\Gamma_Q)^{-1}$ of both the zero-electron and the one-electron
states of the trap may be quite long, and the device may be considered
bistable.

When the voltage $V$ is driven beyond the threshold $V_+$ or $V_-$, the
electron or hole tunnels through the array in time $\tau \sim~C_j/G_j$,
which may be many orders of magnitude shorter than $\tau _L$.  Thus, in
principle, the trap can serve as a memory cell. Its contents can be read
out non-destructively by capacitive coupling of the trap to the
single-electron electrometer\cite {mes,sct,pasct} (see Section
\ref{results}).

Early attempts to trap single electrons were made by Fulton {\it et al.}
\cite{fulton}, using systems with two and four Al/AlO$_x$ junctions of area
$\sim~100\times 100$ nm$^2$ at temperatures down to 0.3 K. Their results
implied trapping times $\tau _L\simeq 1$ sec. Similar experiments by
Lafarge {\it et al.}\cite{lafarge} yielded $\tau _L<1$ sec, much shorter
than could be anticipated from formulas (\ref{gammat}) and
(\ref{gammaq}). A later attempt\cite{nakazato} used a semiconductor
(GaAs) structure with a narrow 2DEG channel instead of a well-defined
tunnel junction array. A bistability loop was observed, but its size was
not clearly quantized, implying that the number of trapped electrons was
much larger than one (the authors estimated this number to be 80-100).

Finally, Al/AlO$_x$ trap circuits designed and fabricated at Stony Brook\cite
{haus,liji} yielded trapping times of over $10^4$ sec (limited only by
observation time). The main goal of the present work was to compare the
experimental data obtained for these traps with a quantitative theoretical
analysis of the circuits. For this purpose, we have constructed a
geometrical model of the circuit, calculated the full matrix of self- and
mutual capacitances for the conducting nodes in the model, and simulated
static and dynamic properties of the trap using these capacitances.

\section{Fabrication}

\label{fabrication} 
Circuits consisting of two layers of partially overlapping nodes were
fabricated using the standard shadow mask technique\cite{dolan,ful-dol}.
The process begins with a Si substrate, either stripped of oxide or covered
by a layer of SiO$_2$ of thickness $H$=500 nm. The substrate is coated with
a PMMA/copolymer double layer mask.  The circuit pattern is written onto
the mask using a scanning electron microscope. Then the mask is developed,
the Al circuit elements are deposited onto the substrate, and the mask is
lifted off.  The fabrication process is described fully in
Ref. \onlinecite{liji}.  Here we present the essential details.

\subsection{Mask}

\begin{figure}
\begin{center}
\leavevmode
\epsfxsize=8cm
\epsffile{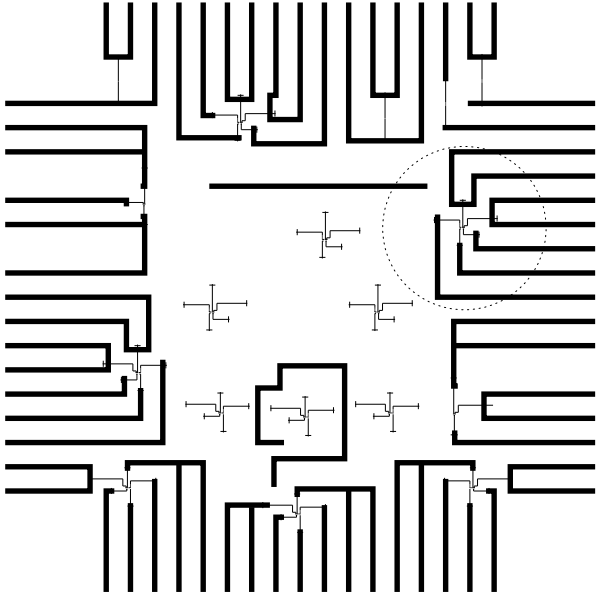}
\end{center}
\caption{Mask for complete chip containing several circuits.  Pattern for
circuits discussed in this paper is circled.}
\label{mask-big}
\end{figure}

\begin{figure}
\begin{center}
\leavevmode
\epsfxsize=8cm
\epsffile{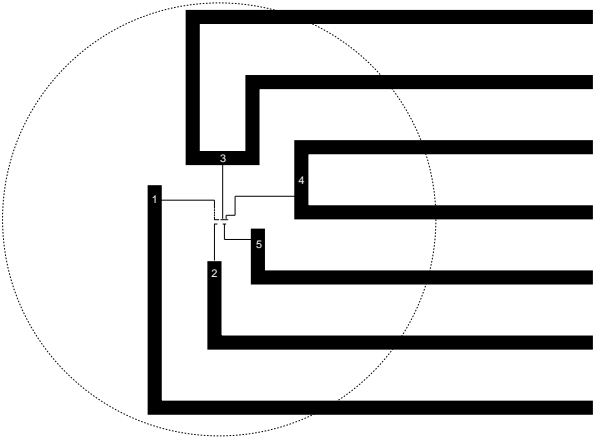}
\end{center}
\caption{Closeup of layout of circuits discussed in this paper, showing 20
$\mu$m cutoff radius used in simulations, and external node numbers.
Externals have wide ($W = 1 \mu$m) and narrow ($w \simeq$ 50 nm) sections.}
\label{mask-med}
\end{figure}

\begin{figure}
\begin{center}
\leavevmode
\epsfxsize=8cm
\epsffile{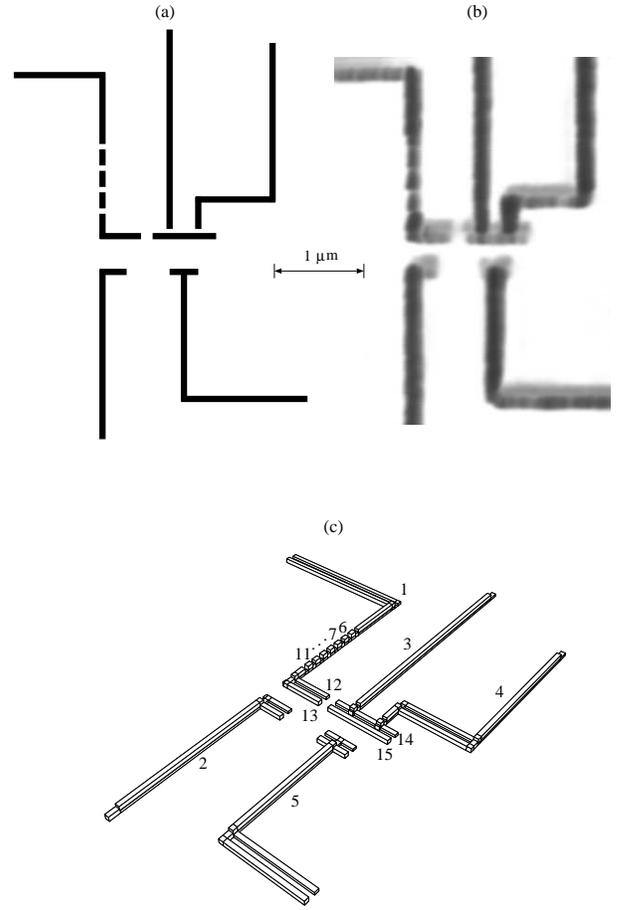}
\end{center}
\caption{(a) Central part of mask. (b) AFM image of central part of
fabricated circuit. (c) 3-D outline of central part of geometrical model
used for capacitance calculation.}
\label{layout}
\end{figure}

The circuit layout, consisting of a set of line segments, is first
specified in a ``mask file'' (Fig.~\ref{mask-big}).  A version of the same
file is also used to start the computational modeling process (see Section
\ref{geometric}).  Wide lines in Fig.~\ref{mask-big} represent the parts of
the externals that extend from the trap and electrometer to contact pads at
the edge of the chip.  The narrow lines extending inward from the wide
lines (Fig.~\ref{mask-med}) represent the inner parts of the externals.
The short, narrow line segments (Fig.~\ref{layout}a) represent islands.

\begin{figure}
\begin{center}
\leavevmode
\epsfxsize=8cm
\epsffile{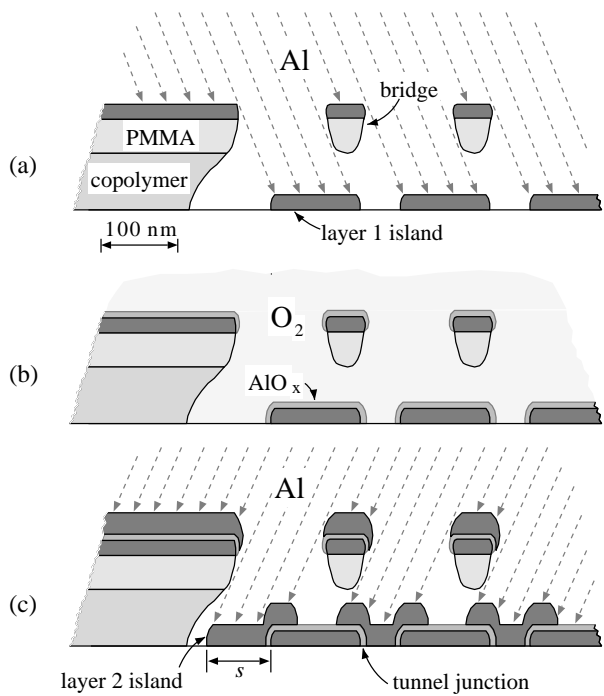}
\end{center}
\caption{Schematic view of shadow mask fabrication: (a) first-layer
evaporation, (b) oxidation (oxide thickness exaggerated for illustration
purposes), (c) second layer evaporation.}
\label{shadow}
\end{figure}

The pattern of lines is written on the mask using the electron beam.  Upon
chemical development, each line in the PMMA becomes a window opening into a
larger cavity in the copolymer, which is more susceptible to the electrons.
This procedure results in a mask, shown schematically in Fig.~\ref{shadow},
with several suspended bridges.

\subsection{Deposition}

The aluminum islands are deposited in two layers. The first layer is
resistively evaporated in high vacuum directly onto the room-temperature
substrate (Fig.~\ref{shadow}a). This layer is then oxidized at $\sim$~10
mTorr O$_2$ for $\sim$~10 min (Fig.~\ref{shadow}b), covering the Al islands
with a $\sim~1$ nm layer of AlO$_x$.  Before depositing the second layer,
the chamber is re-evacuated and the substrate is tilted relative to the Al
source. The tilt creates a shift $s$ between these two groups of islands,
so that they partially overlap (Fig.~\ref{shadow}c).  The AlO$_x$ creates
tunnel barriers between the first and second layer islands.  In our
circuits, the shift $s$ was about 120 nm along the vertical direction in
Fig.~\ref{layout}a,b. The second aluminum layer is made thicker than the
first, to allow reliable step coverage.

An AFM image of the resulting circuit is shown in Fig.~\ref{layout}b. This
image exaggerates the island widths because of the finite angle of the AFM
tip. Other observations (including SEM imaging) show that the islands
oriented perpendicular to the direction of the shift were in fact spatially
separated, in the successful samples.

Figure~\ref{layout}c shows a simplified model of the central part of the
circuit, with externals and islands numbered.  There are two islands for
each corresponding window in the mask.  For example, islands 6 and 7 are
the first- and second-layer products of the same window (see also
Fig.~\ref{model}c.).  Since the two layers of each external overlap each
other extensively and are connected to the same voltage/current source,
they effectively serve as one conductor.  Thus, there is only one external
for each corresponding window in the mask.

\section{Geometrical Modeling}

\label{geometric} The essential electrostatics of a group of conductors can
be described by their mutual capacitance matrix, {\boldmath $C$}. A program
known as FastCap\cite{fastcap} can calculate {\boldmath $C$} for an
arbitrary collection of conductors, given the geometry of the conductors as
input. The conductor surfaces are presented to FastCap as a set of discrete
elements, or ``panels''. We wrote a program called Conpan (for {\it
conductor panels}) to generate a 3D paneling of a simplified model of the
experimental system, starting from a 2D mask file.  We will first explain
the Conpan algorithm, then how its input parameters were derived from
experiments.

\subsection{Conpan Algorithm}

Conpan represents circuit nodes by means of data structures called
``sections''.  Each section is a collection of data about a node or part of
a node.  The data include parameters such as node number, layer number, and
limits in the $xy$ plane.  Sections may be recursively divided into
subsections to represent overlaps and to facilitate paneling.

\begin{figure}
\begin{center}
\leavevmode
\epsfxsize=8cm
\epsffile{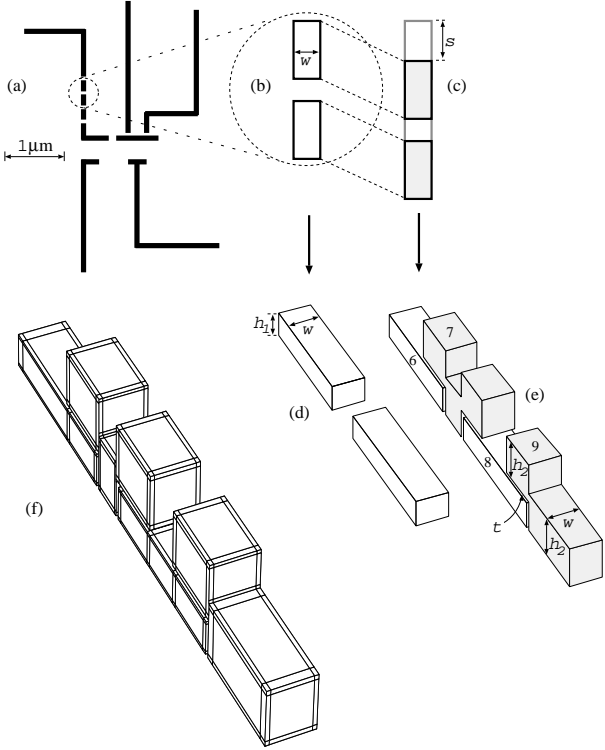}
\end{center}
\caption{Geometrical model construction in Conpan: (a) Two segments of the
array. (b) First-layer sections generated by Conpan for the two segments.
(c) Second-layer island sections (shaded), partially overlapping
first-layer sections. (d) Shape of first-layer islands. (e) Shape of
second-layer islands overlapping first-layer. (f) First- and second-layer
islands paneled for capacitance calculation.}
\label{model}
\end{figure}

Consider two line segments from the larger mask file (Fig.~\ref{model}a).
These two segments eventually produce four islands separated by three
tunnel junctions.  Conpan expands each segment into a first-layer section
(Fig.~\ref{model}b) using the line-width $w$.  The second-layer sections
(Fig.~\ref{model}c) are initially identical to the first-layer sections
except for a uniform translation $s$ that results in overlaps.

\subsubsection{Overlap Detection}

Since a single second-layer section can overlap more than one first-layer
section, Conpan detects the overlaps using a recursive detection algorithm.
To begin, each second layer section is compared against each first layer
section to detect overlaps. When an overlap is found, the second layer
section spawns two daughter sections, one overlapping and one not.  The
axis and coordinate of the split are stored in the the mother section,
along with pointers to the daughter sections.  The mother section 
becomes a placeholder, used only to keep track of the relationship among
its daughter sections.

\begin{figure}
\begin{center}
\leavevmode
\epsfxsize=8cm
\epsffile{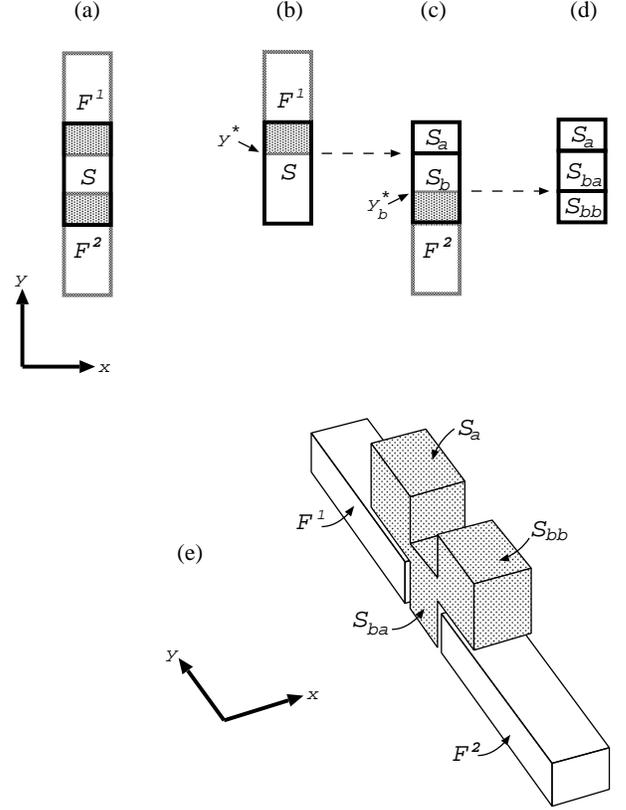}
\end{center}
\caption{Recursive overlap detection in Conpan for a circuit fragment where
a second-layer island $S$ overlaps two first-layer islands $F^1$ and $F^2$.
(a) Top view, with overlap areas shaded.  (b) Comparing $S$ to all
first-layer islands, overlap of $F^1$ is detected.  $S$ is split into two
daughter sections, $S_a$ and $S_b$, at $y = y^*$.  (c) Comparing $S_b$ to
all first-layer islands, overlap of $F^2$ is detected.  $S_b$ is split into
$S_{ba}$ and $S_{bb}$ at $y = y^*_b$. (d) Daughter sections $S_a$, $S_{ba}$,
and $S_{bb}$, having no further overlaps, are used to build the 3D
structure shown in (e).}
\label{divide}
\end{figure}

The non-overlapping daughter is then is compared against the remaining
first-layer sections to find other overlaps.  If there are more overlaps,
the daughter spawns a pair of sub-daughters, and so on.  The recursive
process stops when no new overlaps are found.  The daughter sections that
remain undivided are called ``final daughters''.  Figure~\ref{divide}
gives a schematic view of the recursive overlap detection process for a
single island.

In addition to dividing up the second-layer sections to account for
overlaps, Conpan also splits first-layer sections along the line where they
are overlapped (see Fig.~\ref{model}f).  This allows the edges of panels
facing each other across a junction to line up, facilitating convergence
in capacitance calculations.

\subsubsection{3D Representation}

Once all the overlaps have been found, Conpan can begin to create the 3-D
model of the circuit.  Each final daughter section becomes the base
of a ``block'', a rectilinear solid representing part of a conductor.  The
heights of the two layers are specified by the parameters $h_1$ and $h_2$.
First-layer blocks and non-overlapping second-layer blocks have their base
at $z = 0$.  Overlapping second-layer blocks have their base at $z = h_1 +
t$, where $t$ is the thickness of the gap between overlapping islands that
represents the tunnel junction.

Finally, each block surface is divided into panels.  The goal is to divide
the surfaces in such a way that an acceptably accurate capacitance
calculation can be performed, within the limits of available computer
memory and calculation time.  The division process is guided by an input
parameter $a$, the goal panel length.  The surface of a block with length
$L_i$ along axis $i$ is divided into the number $n_i$ of divisions that
brings $L_i/n_i$ closest to $a$.  Once the block surface has been divided
along both its axes, the resulting grid of panels is written to a panel
file for input to FastCap.  Each panel is stored simply as a quartet of
${x,y,z}$ coordinates, one for each of the four corners, together with the
number of the node it belongs to.

\subsection{Conpan Input Parameters}

\subsubsection{Junction Thickness}

In the physical circuit, the tunnel barriers separating the islands consist
of AlO$_x$, with unknown $x$ and thickness $t_j$. From literature data on
similar junctions \cite{maezawa}, we expect a dielectric constant $\epsilon
_j \sim~4$ and $t_j \sim~1$ nm. FastCap can handle dielectric surfaces much
as it handles conductors -- by dividing them into panels.  However, each
additional dielectric panel demands more computer memory and calculation
time. Since $t_j$ is much smaller than the transversal dimensions in all
junctions, the electric field configuration outside the junctions does not
depend strongly on their internal geometry.  Therefore, we avoided modeling
the junction dielectrics explicitly by replacing them with uniform
free-space gaps ($\epsilon =1$) with the effective thickness
$t=t_j/\epsilon _j$. This effective thickness was adjusted to make the
junction specific capacitance match the standard experimental value
$4.5\mu$F/cm$^2$ typical for the Al/AlO$_x$/Al junctions with tunnel
conductivity in our range ($\sim~10^5$ S/cm$^2$)\cite{maezawa,mager}.

\subsubsection{Line Widths}

\begin{figure}
\begin{center}
\leavevmode
\epsfxsize=8cm
\epsffile{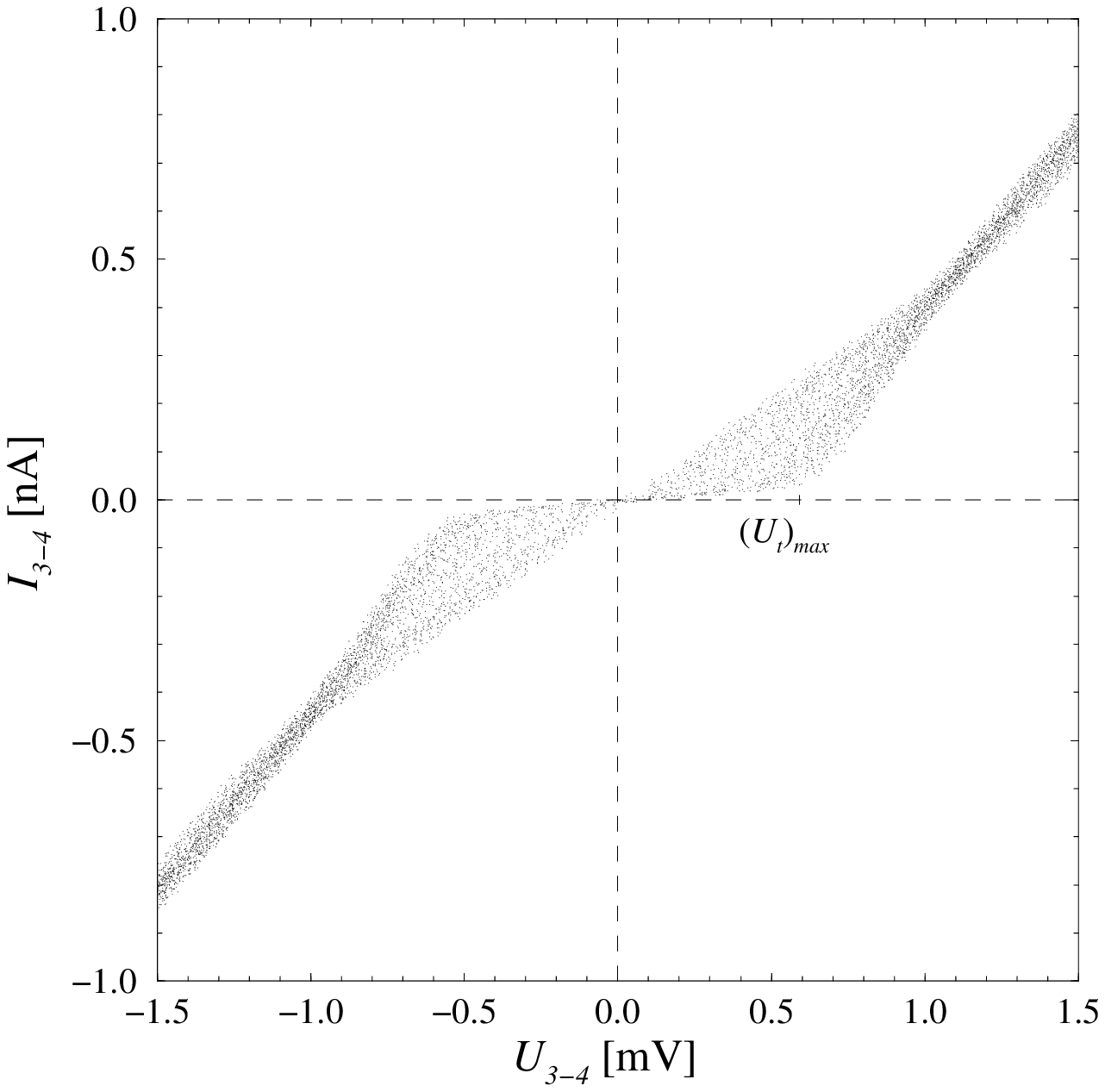}
\end{center}
\caption{Current-voltage traces for the electrometer, (SiO$_2$/Si
substrate, T=35mK, superconductivity suppressed by magnetic field), current
biased, taken while rapidly varying $V_5$.  The Coulomb blockade voltage
$U_t$ varies with $V_5$ from 0 to a maximum value of e$C^{-1}_{14,14}$.}
\label{elect_i-u}
\end{figure}

The effective line width $w$ of islands (and of the narrow parts of
externals) is difficult to measure directly, because of its small magnitude
(see Fig.~\ref{layout}b).  We determined $w$ by requiring that the
simulated inverse self-capacitance of the electrometer island
($C^{-1}_{14,14}$) match its experimentally measured value.  We derive
$C^{-1}_{14,14}$ from the maximum value of the electrometer Coulomb
blockade threshold voltage $U_t$, as seen in electrometer I-V plots
(Fig.~\ref {elect_i-u}):

\begin{equation}
\label{c_sigma}(U_t)_{max} = eC^{-1}_{14,14} \simeq e/C_{14,14}\,. 
\end{equation}

$C_{14,14}$ (conventionally known as $C_\Sigma$) in turn depends on $w$
because it is is dominated by the electrometer junction capacitances, which
increase monotonically with $w$.  The experimentally measured values for
$C^{-1}_{14,14}$ were $(4.6\times 10^{-16}{\rm F})^{-1}$ for the circuit on
Si (sample \#LJS011494B) and $(2.7\times 10^{-16}{\rm F})^{-1}$ for the
circuit on SiO$_2$/Si substrate (sample \#LJS011494A).  The island width $w$
used in simulation, as determined from Eq. (\ref{c_sigma}), was 30 nm for
Si and 42 nm for SiO$_2$/Si. Both of these values are consistent with the
values expected from fabrication parameters and from AFM and SEM imaging of
the samples.

$W$, the width of the wide parts of the externals, is specified as 1 $\mu$m
in our mask files, and can be accepted at ``face value'' because it is
large compared to the scale of geometrical uncertainty in the circuit, and
because the wide parts of the externals are all far (several $\mu$m) from
the islands.

\subsubsection{Layer Heights}

The heights of the two layers ($h_1$ and $h_2$) are determined with a
quartz monitor in the deposition unit during fabrication.  In our case,
these heights were measured to be 30 and 50 nm ($\pm$ 10\%), respectively.

\subsection{Substrate}

To calculate the effects of the substrate on circuit capacitances, FastCap
requires a paneling of the complementary image of the ``footprint'' of the
nodes, because panels representing the dielectric/metal interface (parts of
the substrate covered by nodes) must be treated differently than panels
representing the dielectric/air interface (the exposed substrate). In a
manner analogous to that used for conductor panels (see
Section~\ref{matrices}), one could investigate various methods of paneling
the complementary substrate image in order to minimize the number of
panels, while yet retaining an acceptable level of accuracy. Such a
paneling algorithm itself is not simple to create.

We avoided this problem through an old calculational trick in electrostatics
-- the image method. A modified version of FastCap was created, called
ImageCap, which can simulate the effects of a single- or double-layer
substrate by creating a set (or multiple sets, in the double-layer case) of
image panels.

\begin{figure}
\begin{center}
\leavevmode
\epsfxsize=8cm
\epsffile{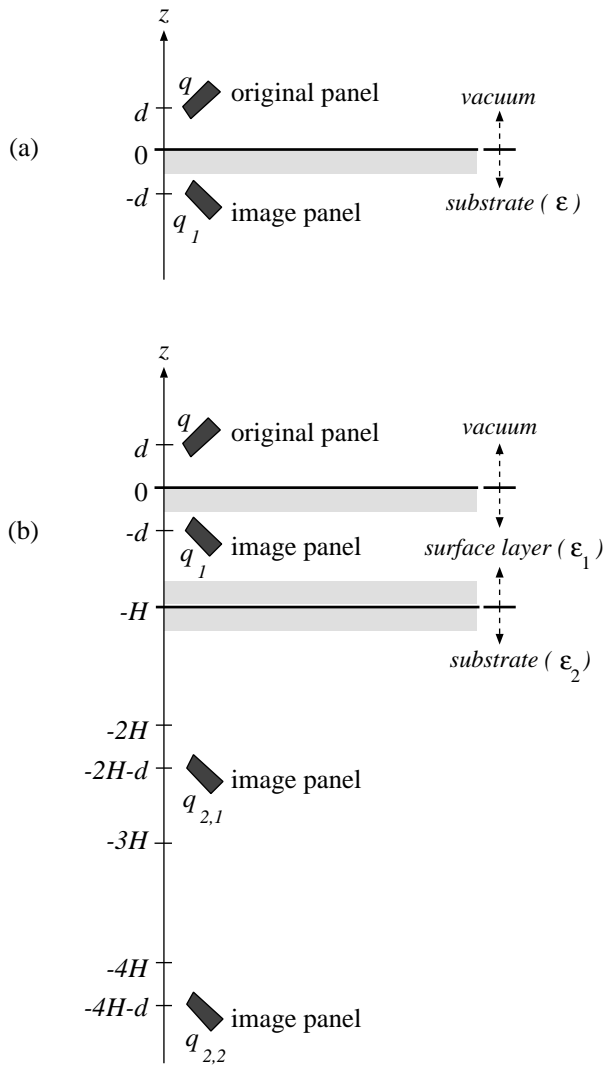}
\end{center}
\caption{Image panels in ImageCap: (a) single substrate, (b) double
substrate.}
\label{image}
\end{figure}

In the single substrate case, each image panel is formed by reflecting the
original panel about the plane of the surface of the substrate
(Fig.~\ref{image}a). For the purposes of calculating the electrostatic
potential above the substrate, the charge on the image panel is
\begin{equation}
\label{image1q}q_1=-\frac{\epsilon -1}{\epsilon +1}q\,, 
\end{equation}
where $q$ is the charge on the original panel, and $\epsilon $ is the
relative dielectric constant of the substrate.

For a substrate covered by an oxide of thickness $H$, an infinite series of
image charges is required for an exact representation of the electrostatic
effect of the substrate (Fig.~\ref{image}b). However, the distance from the
original charge to each successive image charge increases linearly,
\begin{equation}
z_{2,i}=-2Ti-d\,,\quad i=1,2... 
\end{equation}
\noindent 
while the value of each successive image charge decreases exponentially, 
\begin{equation}
q_{2,i}=4\beta \frac{\epsilon _1\epsilon _2}{(\epsilon _1+\epsilon _2)^2}
(\alpha \beta )^{(i-1)}q\,,\quad i=1,2... 
\end{equation}
\begin{equation}
\alpha =\frac{\epsilon _1-1}{\epsilon _1+1}\,,\quad\beta =\frac{\epsilon
_1-\epsilon _2}{\epsilon _1+\epsilon _2}\,. 
\end{equation}
Here $\epsilon _1$ and $\epsilon _2$ are the dielectric constants of the
surface oxide layer and the bulk substrate, respectively. For our circuits,
we have accepted the table values $\epsilon =12.1$ for the bare Si
substrate and $\epsilon _1=4.5$, $\epsilon _2=12.1$ for the SiO$_2$/Si
substrate ($\alpha =0.64$, $\beta =-0.46)$. The resulting expression for
the double-layer image charges,

\begin{equation}
q_{2,i}= -0.36 \times (0.29)^{(i-1)} q \,,\quad i=1,2... 
\end{equation}

\noindent
shows that $q_{2,4}$ is already down by three orders of magnitude from the
original charge. In our calculations, adding image levels beyond $q_{2,4}$
made no difference to the result, within a relative error (of the largest
self-capacitances) below $\sim~10^{-4}$ .

\section{Capacitance Matrices}
\label{matrices}

\subsection{Matrix Structure}

Using the circuit panels generated by Conpan, ImageCap generates the
capacitance matrix for the circuit.  ImageCap adds the effect of image
panels when calculating potentials, and uses no multipole acceleration;
otherwise, its algorithms are the same as in FastCap\cite{fastcap}.  First,
the inverse capacitance matrix for {\it panels} is calculated and inverted.
Each element $\widehat C_{ij}$ in the capacitance matrix for {\it nodes} is
then formed by summing all the panel capacitance matrix elements linking
nodes $i$ and $j$.  The charges and potentials on the nodes are related by

\begin{equation}
\vec q = \mbox{\boldmath $\widehat C$} \vec \phi\,, 
\label{mat_eq}
\end{equation}

\noindent
so that $\widehat C_{ij}$ is numerically equal to the amount of charge
induced on node $j$ when node $i$ is held at unit potential and all other
nodes have zero potential.

{\boldmath $\widehat C$} is an $N \times N$ matrix, where $N = N_e + N_i$,
and $N_e$ and $N_i$ are the numbers of external nodes and island nodes in
the circuit, respectively.  Ordering all the external nodes before the
island nodes, we can write {\boldmath $\widehat C$} in terms of
submatrices:

\begin{equation}
\mbox{\boldmath $\widehat C$}=\left(
\begin{array}{c|c}
\bigotimes&\mbox{\boldmath $-\tilde C$}\\
\hline
\mbox{\boldmath $-\tilde C^T$}&\mbox{\boldmath $C$}\\
\end{array}\right)\,.
\label{quad_mat}
\end{equation}

\noindent
Here {\boldmath $C$} is the symmetric $N_i \times N_i$ matrix of
island-island capacitances and {\boldmath $\tilde C$} is the $N_e \times
N_i$ matrix of external-island capacitances (with elements defined positive,
by convention).  External-external capacitances (represented above by the
$\bigotimes$) are not needed for our simulations.

The matrices calculated for our circuits are shown in Tables I and II.
Note the up/down alternation of mutual capacitances along the array for the
circuit on Si -- e.g., $\tilde C_{1,i}$, the capacitances linking
external node 1 to the islands.  For example, although island 7 is closer
to external 1 than is island 8, $\tilde C_{1,7} \simeq 0.030 \times
10^{-16}$F is smaller than $\tilde C_{1,8} \simeq 0.045 \times 10^{-16}$F
(similarly for islands 9 and 10).  In {\boldmath $C$}, we see that
$\|C_{6,9}\|$ is smaller than $\|C_{6,10}\|$, etc.  This phenomenon
reflects the influence of the silicon substrate, which, due to its high
dielectric constant ($\epsilon \simeq 12$), links externals to the
first-layer islands (which lie flat on the substrate) more strongly than to
the second-layer islands (which lie partly on top of the first-layer
islands).  The capacitances for the circuit on SiO$_2$/Si do not show these
oscillations as strongly, as we would expect from the smaller permittivity
($\sim~4.5$) of SiO$_2$.

\subsection{Model Accuracy}
Our model contains three main simplifications related to computational
constraints, each of which introduces error into our capacitance matrix
calculations.

\subsubsection{Free-space Junctions}

As noted above, we calculate {\boldmath $\widehat C$} using free-space
junctions of thickness $t$ instead of dielectric junctions of thickness
$t_j$.  (Although initially this approximation was intended for
convenience, it later became a necessity as ImageCap does not handle
explicit dielectric panels.)  The error involved in this approximation was
estimated by using FastCap to model a chain of islands in two ways: with
explicit dielectric junctions and with free-space junctions.  Results for
an 8-island chain, with effective dielectric thickness chosen to make the
island self-capacitances in both models the same, indicate that the error
involved in this approximation is below 1\% for junction-linked islands,
and between 1\% and 4\% for non-junction-linked islands.

\subsubsection{Paneling}

\begin{figure}
\begin{center}
\leavevmode
\epsfxsize=8cm
\epsffile{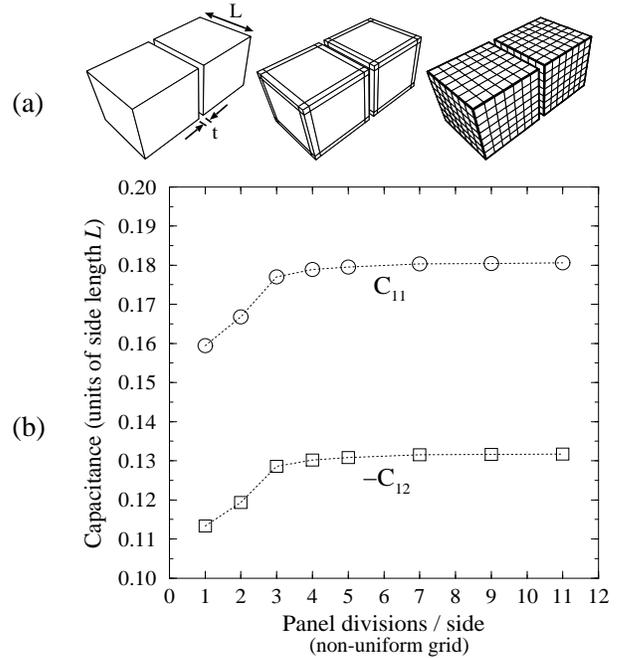}
\end{center}
\caption{Two cube system with $L/t=10$: (a) Panelings with 1, 3, and 9
panels/side. (b) Capacitances as a function of panel density ($\vec q =
\mbox{\boldmath $C$} \vec \phi$).}
\label{2cube}
\end{figure}

In calculating capacitances, FastCap/ImageCap assigns a uniform charge
distribution to each panel.  Hence, its accuracy depends on how well the
paneling follows changes in change distribution on the node surfaces.
Clearly, the denser the paneling, the better the representation of changes
in charge distribution.  However, panel density is effectively limited by
available computer memory.  For example, a FastCap simulation with 5000
panels typically requires more than 128 MB.  ImageCap uses even more
memory, since it calculates all panel interactions directly.  We
investigated the dependence of calculated capacitance on paneling density
for a simple two cube system (Fig.~\ref{2cube}a). 

\begin{figure}
\begin{center}
\leavevmode
\epsfxsize=7cm
\epsffile{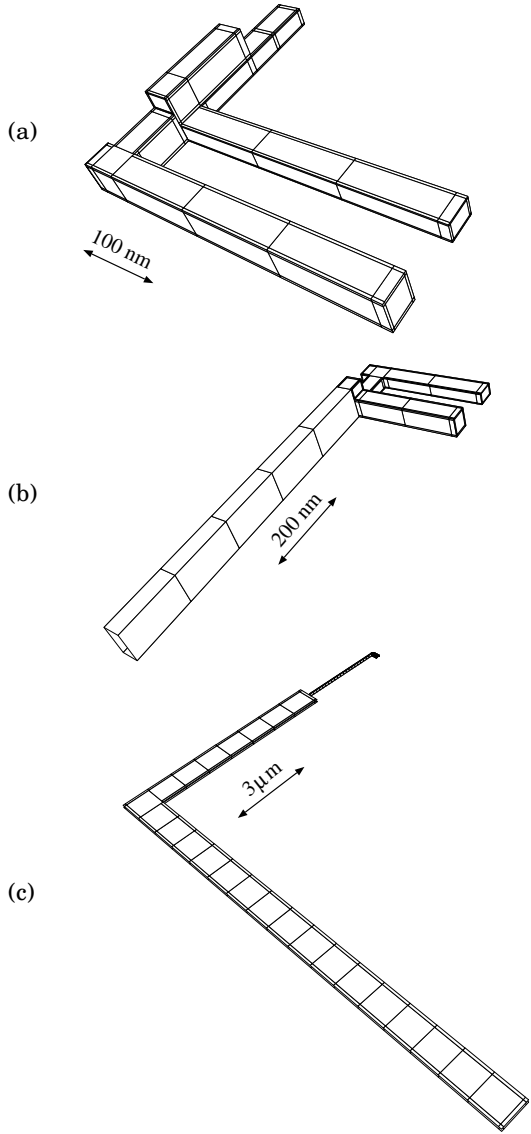}
\end{center}
\caption{Paneling used in our calculations: (a) Trap islands (nodes 12 and
13). (b) Thin parts of externals. (c) Wide parts of externals.}
\label{paneling}
\end{figure}

The results (Fig.~\ref{2cube}b) suggest that a non-uniform $3 \times 3$
grid (with a 1/10 ratio of edge panel length to central panel length,
reflecting the peak in surface charge near the edges) for the smaller,
roughly square-shaped node faces (Fig.~\ref {model}f) is sufficient to
calculate capacitances with an error below 10\%.  This is essentially how
we paneled roughly square-shaped island surfaces.  For longer faces
(Fig.~\ref{paneling}a) we used a larger number of divisions along their
length.  In an islands-only test circuit, increasing the total number of
panels from $\sim 2000$ (corresponding to the $3 \times 3$ grid for roughly
square-shaped surfaces) to 6000 resulted in less than 1\% changes in
island-island capacitances.  Thus we believe that the total error in
island-island capacitances due to finite panel density is perhaps only
$\sim$~1\%.

\begin{figure}
\begin{center}
\leavevmode
\epsfxsize=8cm
\epsffile{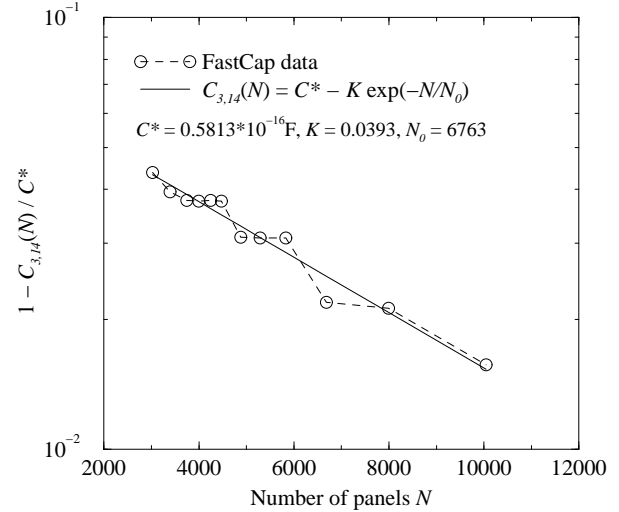}
\end{center}
\caption{Effect of panel density on the calculated island-external
capacitance.  The test circuit had only islands 14 and 15 and narrow parts
of externals.}
\label{c-ext}
\end{figure}

To reduce the number of panels in the model, the two layers of an external
are fused into one where they overlap.  The error involved in this
simplification is negligible.  In addition, the narrow parts of the
externals were divided along their length without edge panels
(Fig.~\ref{paneling}b).  This simplification was found to cause an error in
island-external capacitance of $\sim$~5\% when the island and the external
are connected by a junction (Fig.~\ref{c-ext}), and $\sim$~1\% otherwise.
Finally, wide parts of the external leads were represented by only their
top and bottom surfaces (Fig.~\ref{paneling}c), again to save panels.
Since the width to height ratio $W/h \simeq 12$, the error introduced by
this simplification is negligible.  The top and bottom surfaces are divided
according to the $3 \times 3$ type scheme described above for islands.
Despite the large size of the resulting panels, the error involved in this
simple paneling appears to be $\sim$~1\%.

\subsubsection{Lengths of Externals}
\label{lengths}
\begin{figure}
\begin{center}
\leavevmode
\epsfxsize=8cm
\epsffile{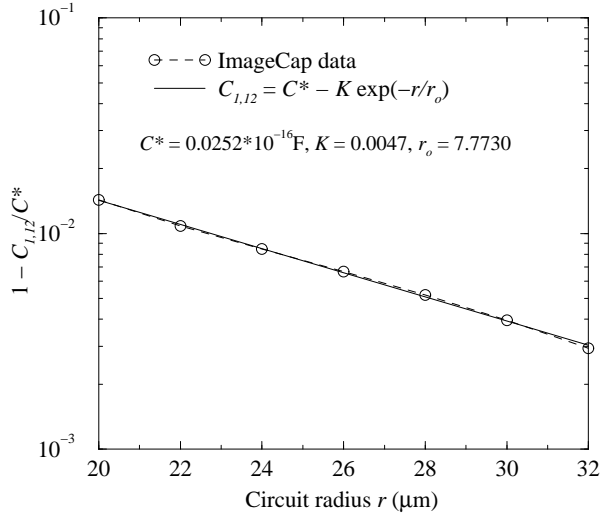}
\end{center}
\caption{Effect of lead length cutoff on the calculated island-external
capacitance.  The test circuit had only island 12 and wide parts of
externals.}
\label{c-wire}
\end{figure}

The calculated capacitance values depend on the lengths of the external
wires used in the model.  In general, island-external capacitances increase
with external length, at the expense of island stray capacitance
(capacitance to a ground at infinity); the self-capacitance of islands does
not change appreciably.  To measure the error introduced by cutting off the
externals at a given length, we have calculated capacitance matrices for test
circuits with varying external lengths (Fig.~\ref{c-wire}).  These circuits
consisted of only one island and only the wide parts of the five externals.
As a result, the error induced by cutting off externals in these test
circuits should be proportionately larger than the error in the complete
circuits.  Still, the test circuits indicate that the error involved in
cutting of the circuit at a radius of 20 $\mu$m (as in our final versions
of the complete circuits) was less than 2\%.

\subsubsection{Total Error}
Considering the error caused by the above simplifications in the
calculation of {\boldmath $\widehat C$} itself, it seems safe to say that
the the combined error for any given calculated capacitance matrix element
was less than 10\%.  Note that we are not yet considering how well the
geometrical model corresponds to the physical circuit (see
Section~\ref{discussion}).

\section{Simulated and Experimental Results}
\label{results}
We have calculated most properties of our circuits using {\sc moses}, the
single-electron circuit simulation program\cite{moses}. This program uses a
Monte Carlo algorithm to simulate arbitrary SET circuits within the
framework of the orthodox theory of single-electron tunneling
\cite{mes,sct}.  {\sc moses} needs to know the capacitance sub-matrices
{\boldmath $C$} and {\boldmath $\tilde C$} and the conductances of all
tunnel junctions. The resistance of two electrometer junctions connected in
series can be extracted from the slope of the experimental dc $I-V$ curve
of the electrometer at high voltage ($V\gg e/C_\Sigma$). From this
measurement, we calculated tunnel conductance per unit area.  Conductances
of all other junctions in the circuit were then calculated by assuming that
their conductance is proportional to their nominal area.  This
assumption may only be accurate to an order of magnitude; however, most of
the results discussed below pertain to stationary properties of the system,
and are thus unaffected by deviations in conductance.

\subsection{General Electrostatic Relations}

Solving the matrix equation (\ref{mat_eq}) for the island potentials
$\phi_i$, with our definition (\ref{quad_mat}) of the capacitance matrix we
get

\begin{equation}
\phi_i=\sum_{j\in isl}{C^{-1}_{ij}(q_j + \tilde q_j)}\,,\quad
\tilde q_j \equiv \sum_{k\in ext}{\tilde C_{kj}V_k}\,,
\end{equation}

\noindent
or, in a different form,

\begin{equation}
\phi_i=\sum_{j\in isl}{C^{-1}_{ij}q_j}+\sum_{k\in ext}{\alpha_{ik}V_k}\,,\quad
\alpha_{ik} \equiv \sum_{j\in isl}{C^{-1}_{ij}\tilde C_{kj} }\,,
\label{phi_i_def}
\end{equation}

\noindent
where $V_k$ are the external potentials.  These relations allow us to
establish useful relations between changes in the external potentials
$\{V_k\}$ and the charge state of the islands $\{q_i\}$, and the dynamics
of the system as determined by the island potentials $\{\phi_i\}$.

\subsection{Electrometer}

Let us apply these relations, in particular, to the island of the
single-electron transistor (number 14 in our notation, see
Fig.~\ref{schematic}) serving as the electrometer.  Experimentally, we
measure the dc voltage $U_{3-4}$ between the ``source'' and ``drain'' of
the transistor (externals 3 and 4) under a small ($\sim$~100 pA) dc
current bias.  If the temperature is small enough ($k_B T \ll
e^2C^{-1}_{14,14}$), the voltage $U_{3-4}$ in such an experiment closely
follows the threshold $U_t$ of the Coulomb blockade of the transistor --
see Fig.~\ref{elect_i-u}.

It is well known (see, e.g., Refs.~\onlinecite{mes,sct}) that the threshold
is determined by the effective background charge $Q_o$ of the transistor
island, which may be defined as

\begin{equation}
\phi_{14}|_{U_3=U_4=0} = C^{-1}_{14,14}(q_{14}+Q_o)\,.
\label{qo_intro}
\end{equation}

\noindent
Comparing (\ref{qo_intro}) and (\ref{phi_i_def}) above, we obtain in our
notation

\begin{equation}
Q_o = \frac{1}{C^{-1}_{14,14}}
\left[
\sum_{j\in isl}^{j\ne 14}{C^{-1}_{14,j}q_j}+\sum_{k\in ext}{\alpha_{ik}V_k}
\right]\,.
\label{qo_sums}
\end{equation}

Eq. (\ref{qo_sums}) allows us to find the theoretically expected variation
of $Q_o$ due to any changes in the system.  On the other hand, the
threshold voltage is an e-periodic function of $Q_o$, and its maximum
amplitude is expressed by Eq. (\ref{c_sigma}) (for the case when the two
transistor junction capacitances are the same).  Thus, after we measure the
experimental value of $(U_t)_{max}$, we can express the change in the
effective charge $Q_o$ via the observed variation in $U_t$:

\begin{equation}
\Delta Q_o = \frac{e\Delta U_t}{2(U_t)_{max}}\,.
\label{qo-ut}
\end{equation}

We have applied this approach to compare experiment and theory for two
samples (\#LJS011494A with SiO$_2$/Si substrate and \#LJS011494B with Si
substrate).

\subsection{High-$T$ electrometer response}

\begin{figure}
\begin{center}
\leavevmode
\epsfxsize=8cm
\epsffile{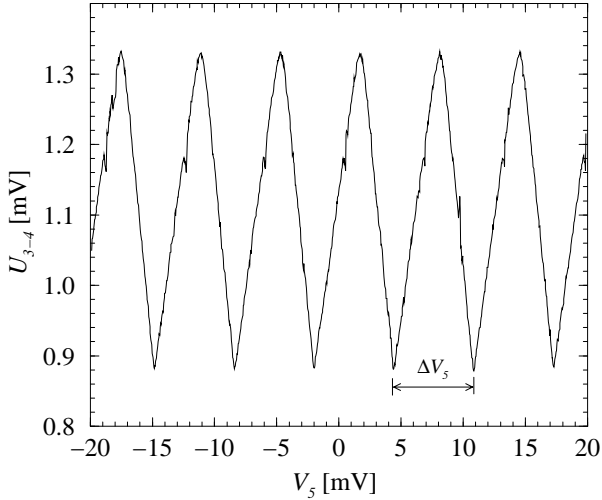}
\end{center}
\caption{A typical experimental dependence of electrometer dc voltage
$U_{3-4}$ on gate voltage $V_5$ for a circuit comprising only an
electrometer.}
\label{elect_u-v5}
\end{figure}

We can readily measure $\Delta V_i$ (i = 1,2,5), the change in external
voltage corresponding to one period of the oscillating threshold voltage
(Fig.~\ref{elect_u-v5}).  At $k_BT\geq 0.1e^2/C_\Sigma$ (experimentally,
$T\geq 0.5K$), thermal activation of electrons smears the Coulomb blockade
effects and makes the junctions essentially transparent to tunneling, while
the periodic response of the electrometer is still visible up to $k_BT \sim
0.3e^2/C_\Sigma$ ($T \sim~1.5$K). Thus, the measured values of $\Delta V_i$
depend only on the circuit geometry and are essentially independent of the
properties of the junctions.

{\sc moses} is not a useful tool for directly modeling high-$T$ behavior,
as the number of jumps involved would be extremely high. However, we can
simulate high-$T$ behavior in {\sc moses} by specifying very high external
voltages $V_i$ while keeping temperature low (say, $T=0$). Under these high
voltage conditions, the islands are flooded with extra electrons, and the
tunnel junctions become effectively transparent to tunneling, just as in
the high temperature case. Thus we simply apply an external voltage $V_i
\gg e/C_\Sigma$, measure how many electrons enter the electrometer island,
and find the ratio of voltage $V_i$ to electrometer charge $q_{14}$.

Table III shows values of the ratio $\Delta V_i$ for simulated circuits and
for experimental circuits averaged over several nominally identical
samples.  For the experimental values, the uncertainties given reflect the
spread of among the samples.  For the simulated values, the uncertainties
given reflect the $\sim$~10\% error in calculated values, as described in
Section~\ref{matrices}.  The simulated values are all lower than the
experimental ones (with the exception of $\Delta V_5$ on SiO$_2$/Si),
differing by as much as 50\%.  The agreement is somewhat better for the
circuits on SiO$_2$/Si.

\subsection{Trap phase diagram}

\begin{figure}
\begin{center}
\leavevmode
\epsfxsize=8cm
\epsffile{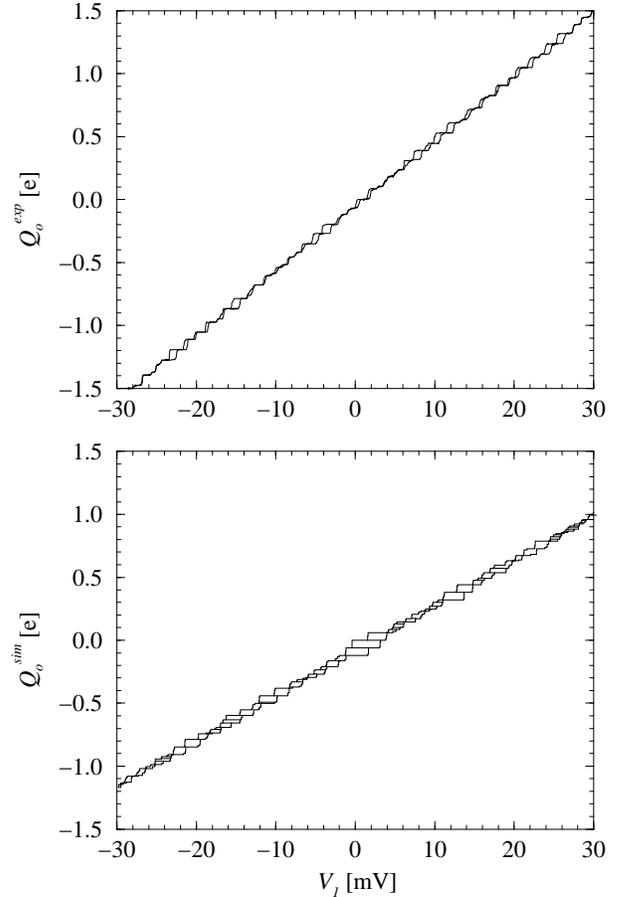}
\end{center}
\caption{Electrometer phase diagram: $Q_0$ as a function of trap drive
voltage. $V_2$=8.2mV, sample on SiO$_2$/Si.  Crosstalk from drive voltage
to electrometer has been subtracted, leaving only influence of trap charge
state.  In experiment, superconductivity in aluminum is suppressed by a 2T
magnetic field.}
\label{phase}
\end{figure}

The simplest measurable characteristic involving single-electron charging
of the trap is its phase diagram (Fig.~\ref{phase}), which reflects changes
in the charge states of the array and trap as a function of the drive
voltage $V_1$. In {\sc moses}, we can directly view the charge state of
each island in the array and trap as we vary $V_1$, as well the resulting
change in $Q_o$.  In the physical circuit, however, we can only measure the
response $U_{3-4}$ of the electrometer and reduce it to the changes in
$Q_o$ using Eq.~\ref{qo-ut}.

Figure ~\ref{phase} shows experimental and simulated electrometer response
to ramping the trap drive voltage $V_1$ up and down over a period of
several minutes.  In both cases, the effects of the crosstalk between
external 1 and the electrometer have been removed.  In the experiment, the
crosstalk is cancelled by feeding the electrometer gate (node 5) with a
voltage $V_5=-\alpha V_1$, with the coefficient $\alpha$ adjusted to make
the phase diagram plateaus horizontal.  In simulations, {\sc moses}
accomplishes the same effect by subtracting $\Delta \phi_{14} =
\alpha_{1,14}V_1$ from the electrometer island potential.

\begin{figure}
\begin{center}
\leavevmode
\epsfxsize=8cm
\epsffile{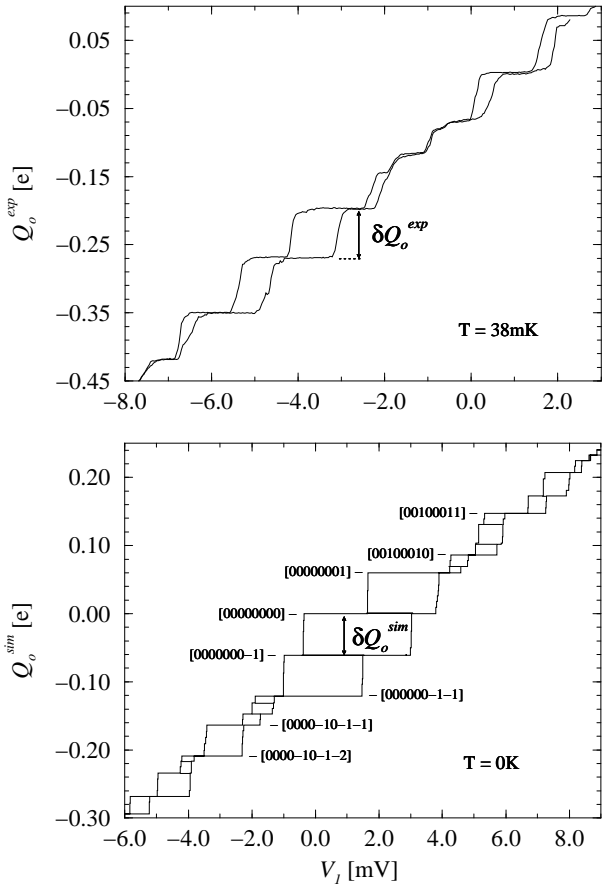}
\end{center}
\caption{Closeup view of experimental and simulated phase loops.  Charge
vectors for islands 6-13 on various plateaus in simulated phase diagram are
indicated in brackets.  In simulation, $\vec q_o = 0$ was assumed.}
\label{loops}
\end{figure}

\begin{figure}
\begin{center}
\leavevmode
\epsfxsize=8cm
\epsffile{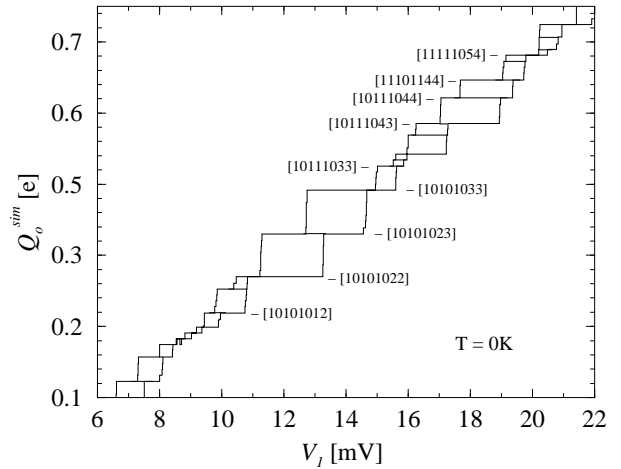}
\end{center}
\caption{Simulated phase loops at a higher range in $V_1$, showing wider
plateaus around [1010123].}
\label{hi-loops}
\end{figure}

Horizontal plateaus in Fig.~\ref{phase} correspond to particular charge
states of the system (trap~+~array), while vertical jumps correspond to
changes of charge state.  Thus, the hysteretic loops are regions of
bi/multi-stability.  The blow-up of the theoretical curve (Fig.~\ref{loops})
indicates the states for several plateaus.  In particular, notice that the
largest plateaus correspond to states that are most stable because the array
is either charged uniformly (one electron on each island, for example) or
in a regular alternating pattern such as 1-0-1-0 (Fig.~\ref{hi-loops}).
The smaller plateaus correspond to more complex charge states which are
less stable.

The experimental phase diagram bears a qualitative resemblance to the
theoretical one, with somewhat shorter plateaus, though the order of
magnitude is the same ($\sim$~2 mV for major plateaus).  Simulated phase
diagrams with randomly selected $\vec q_0$ show shorter plateaus than the
$\vec q_0 = 0$ phase diagram (see Sec.\ref{discussion} below).

In Fig.~\ref{phase}, the large jumps in $Q_o$ correspond to a single
electron entering the trap: $\delta Q_o \equiv (\Delta Q_o)|_{e\to tr}$.
Using Eq.~(\ref{qo_sums}), we can also express the simulated value of
$\delta Q_o$ as

\begin{equation}
\delta Q_o^{sim} = \frac{C^{-1}_{tr,14}}{C^{-1}_{14,14}}e\,,
\end{equation}
\noindent
where $tr$ = 12 or 13, depending on which trap island the electron stops
in.  For comparison with experimental results, we take the average of the
two possible values.  The results are shown in Table IV.  The difference
between simulated and experimental values for Si is within the estimated
geometric calculation error (10\%), while the value for
SiO$_2$/Si is not.

\subsection{Plateau dependence on $V_2$}

\begin{figure}
\begin{center}
\leavevmode
\epsfxsize=8cm
\epsffile{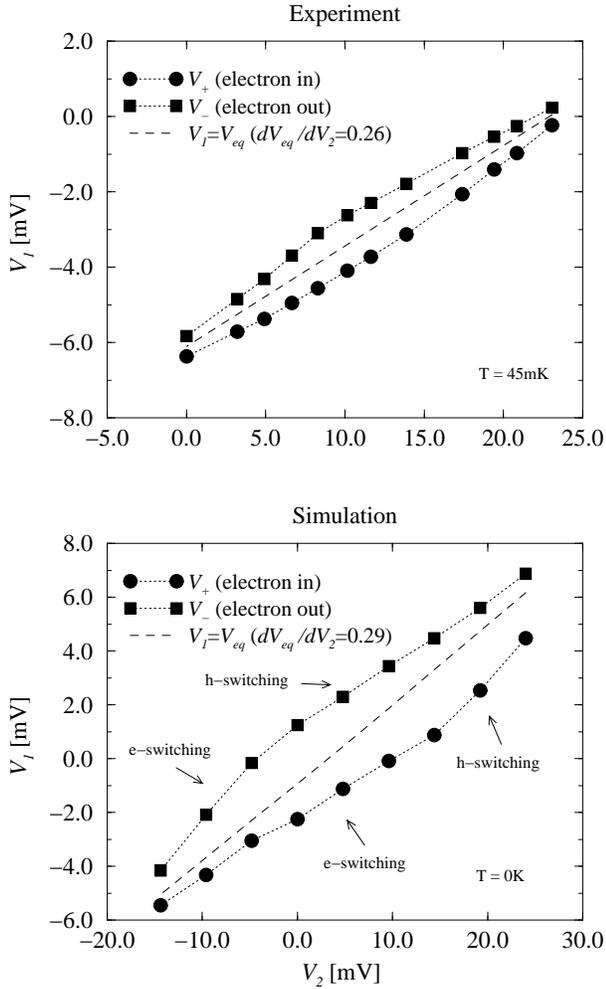}
\end{center}
\caption{Experimental and simulated threshold voltages $V_\pm$ as functions
of $V_2$: Si substrate, superconductivity suppressed.}
\label{diamond}
\end{figure}

For a given plateau, the switching voltages $V_1$ = $V_\pm$ depend on the
``ground'' voltage $V_2$ (see Fig.~\ref{schematic}).  In the simplest
model, with no stray capacitances, (see, e.g., Ref. \onlinecite {pasct})
the charge state of the system depends only on the voltage $V=V_1-V_2$.  In
that model, the dependences $V_\pm(V_2)$, corresponding to changes in the
charge state, would form parallel 45$^{\circ}$ lines in the $[V_1,V_2]$
plane.  In reality, however, stray capacitances of the islands to
``infinity'' (i.e. to a distant common ground) make the average potential
$(V_1+V_2)/2$ of the system relevant as well. As a result, the region
corresponding to each charge state acquires a shape similar to a stretched
diamond (Fig.~\ref{diamond}).

Simulations using {\sc moses} show that the diamond shape results from the
alternation of two types of electron transport that switch the charge
state. At the low-$V_2$ end of the diamond, the charge state switches
with the transfer of an electron in/out of the trap (see the energy diagram
in Fig.~\ref{profile}a).  However, at the high-$V_2$ side, the barrier for
holes to enter or exit is lower than for electrons
(cf. Fig.~\ref{profile}b).  Near the sharp ends of the diamond, the
critical transport may be even more complex (e.g., creation of an
electron-hole pair inside the array, with the sequential motion of its
components apart, one into the trap, and another into the external
electrode). In these regions, however, the plateau corresponding to the
charge state of the trap is already small and virtually disappears among
numerous plateaus corresponding to various internal charge states of the
array (Fig.~\ref{loops}).  Figure~\ref{diamond} shows that while the
diamond shape of the charge state in $[V_1,V_2]$ is well reproduced in
experiment, the simulated width \mbox{$\|V_{+}-V_{-}\|$} of the bistability
region in $V_1$ is roughly twice the experimental value.

For each $V_2$, there is one value of $V_1$, called $V_{eq}$, at which the
energy barrier is the same for an electron to tunnel into or out of the
trap\cite{seneca}.  A good measure of the relative influence of the two
external voltages on the trap is the derivative

\begin{equation}
\frac{dV_{eq}}{dV_2} = \frac{\alpha_{2,tr}}{1 - \alpha_{1,tr}}\,,
\end{equation}

\noindent
where $tr$ = 12 or 13, depending on which trap island actually traps the
electron for a given $(V_1,V_2)$.  The two values are typically within 5\%
of each other, and we take their average when comparing simulated and
experimental results.  In the experimental data, we define the average
$V_{eq}$ by bisecting the diamond shape in the graph (Fig.~\ref{diamond}).
$dV_{eq}/dV_2$ is essentially a geometric property of the circuit and
should not depend on thermal activation or cotunneling.  As
Fig.~\ref{diamond} shows, the simulated and experimental values are very
close.

\subsection{Energy barrier}

\begin{figure}
\begin{center}
\leavevmode
\epsfxsize=8cm
\epsffile{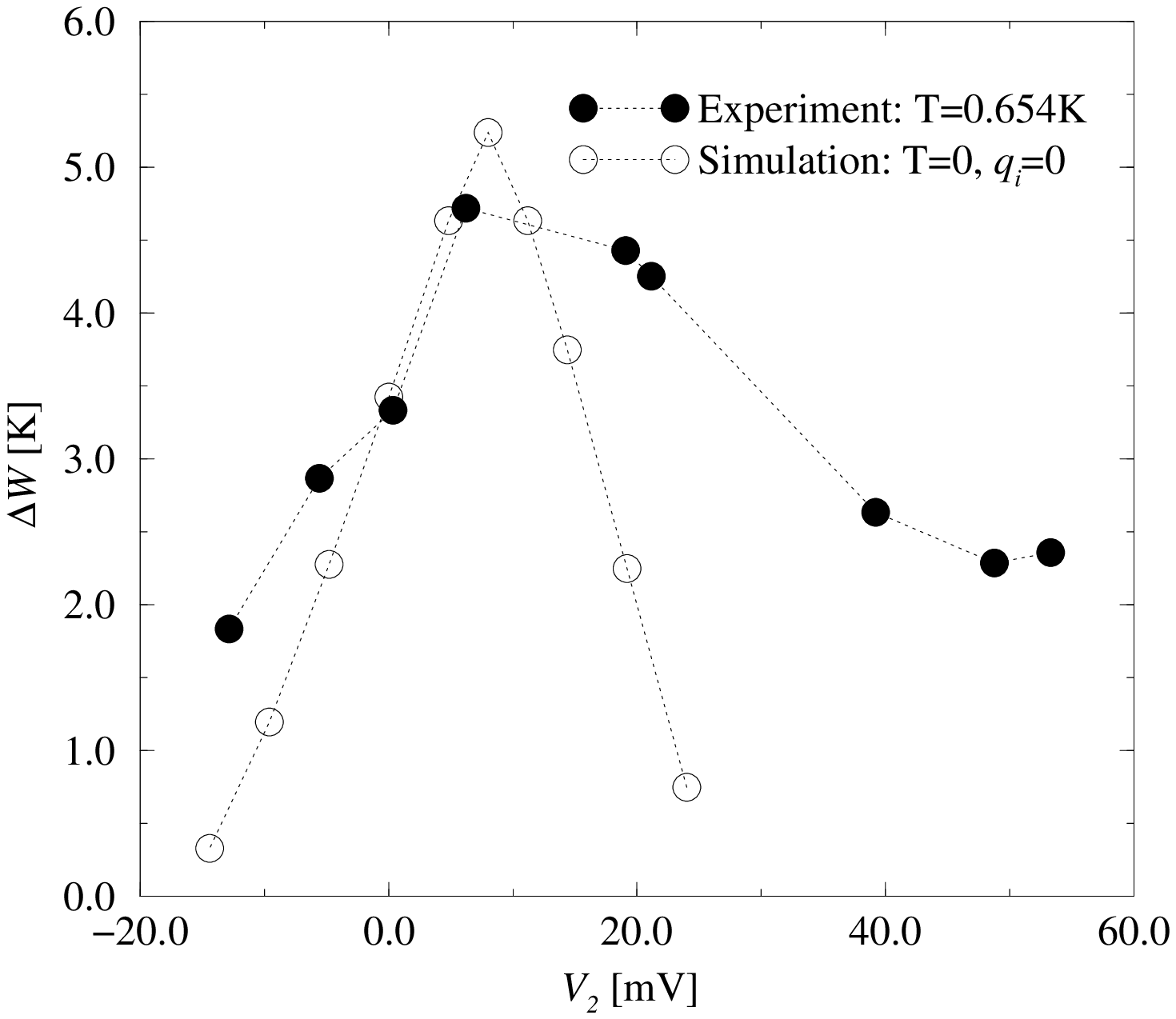}
\end{center}
\caption{Experimental and simulated dependence of barrier height on $V_2$
(SiO$_2$/Si substrate).}
\label{dW_v2}
\end{figure}

At $V_1=V_{eq}(V_2)$, we can measure the energy barrier $\Delta W$
experimentally by measuring trapping lifetime as a function of temperature
(for experimental details, see Ref.~\onlinecite{haus}).  The Arrhenius law
for lifetimes gives

\begin{equation}
\tau_L \propto \exp (\Delta W/kT),
\label{arrhen}
\end{equation}
\noindent
so that plotting $\log (\tau_L)$ vs. $1/T$ gives us $\Delta W$.  Dynamical
simulations\cite{seneca} have shown that (\ref{arrhen}) is virtually
unaffected by cotunneling for relatively high temperatures ($\sim~100$ mK
and above).  In simulation, {\sc moses} allows us to measure $\Delta W$
directly. Figure~\ref{dW_v2} shows the dependence of the trap energy
barrier $\Delta W$ on the bias voltage $V_2$, for the circuit on the
SiO$_2$/Si substrate. The simulated energy barrier profile peaks at roughly
the same value of $V_2$ as in the experiment, and the peak barrier value is
within $\sim~10\%$ of the experimental value.  However, the simulated peak
is sharper.

\section{Discussion}
\label{discussion}
Let us first discuss the results independent of the single-electron
charging effects: the transistor response $\delta Q_o$ to a single electron
entering the trap, the oscillation periods $\Delta V_i$, and the slope
$dV_{eq}/dV_2$.  The differences between simulated and experimental values
for $\delta Q_o$ are 6\% and 23\% for the Si and SiO$_2$/Si substrates,
respectively.  Values for $\Delta V_i$ do not agree as well: differences
between simulated and experimental values range from 15 to 38\% for the
trap on SiO$_2$/Si, and from 29 to 50\% for the trap on Si.  We had
experimental data for $dV_{eq}/dV_2$ only on Si.  Here the difference
between experimental and simulated results was $\sim~12\%$.  These numbers
suggest how well our geometrical model corresponds to the physical circuit
(the accuracy of the orthodox theory and of the {\sc moses} simulator is
presumably much higher).

The most obvious idealization involved in our geometric modeling is that
the islands created by Conpan are rectilinear and uniform. Even at the
limited resolution of an AFM image (Fig.~\ref{layout}b), the contours of
the fabricated circuits appear rounded and irregular on a scale of $\sim$
10 nm. This is to be expected, due to the relatively large grain size of
evaporated Al ($\sim~50$ nm, comparable to the line width $w$) and the
stochastic nature of the grain growth process.  Most capacitance matrix
elements should not depend strongly on small details of the island shape.
However, irregularities in the shape of overlapping islands may change the
area, and thus the capacitance, of the junctions linking them.

All other results involve single-electron charging effects. Here the
difference between the theory and experiment is larger - typically by a
factor of 2, and sometimes larger. We believe that the most important
origin of this difference is the set of background charges $\vec{q}_0$.
The Si substrate is capable of trapping charged impurities near the circuit
islands.  The result of these impurities is that the charge on island $i$
effectively changes from $\tilde{q}_i$ to $\tilde{q}_i + q_{0i}$.  These
charges may furthermore be capable of thermal migration over time.

\begin{figure}
\begin{center}
\leavevmode
\epsfxsize=8cm
\epsffile{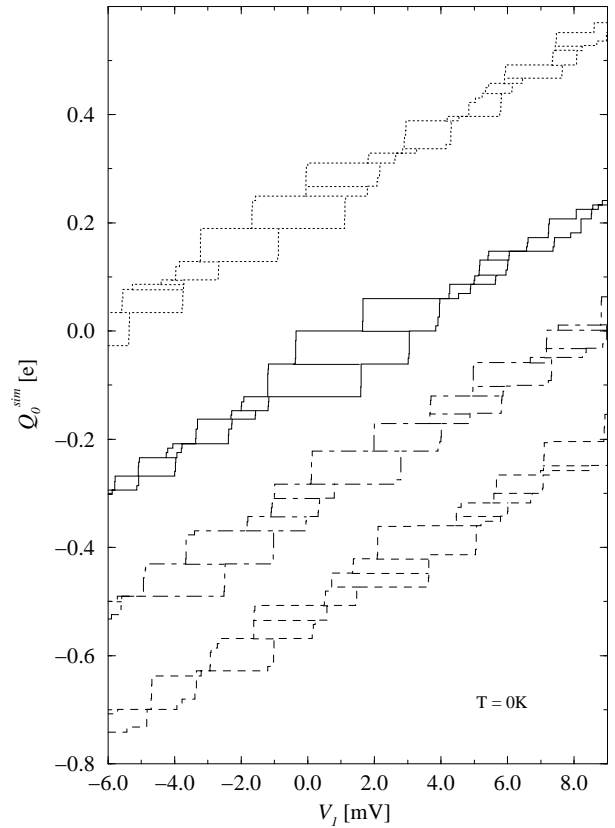}
\end{center}
\caption{Simulated response of electrometer to ramps of $V_1$ for $\vec q_0
= 0$ (solid line) and three randomly selected background charge vectors
$\vec q_0$.  Plots are shifted vertically because background charge on
the electrometer island (node 14) effectively shifts $Q_o$.}
\label{multi_q0}
\end{figure}

Simulated plots of the electrometer response to trap charging with three
randomly selected $\vec q_0$ are shown in Fig.~\ref{multi_q0}.  It appears
that the wide ($\sim$~4 mV) steps near $V_1 = 0$ in the $\vec q_0 = 0$ plot
are not stable to variations in $\vec q_0$: in most plots with random $\vec
q_0$, as in the experimental plot, all step widths are less than 3 mV.

To summarize: we have developed an automated way to construct simplified
models of experimental single-electron devices and circuits with metallic
islands, and we have compared the properties of model single-electron traps
with those of real traps.  The observed differences between simulation and
experiment may be attributed to random deviations of the physical
structures from their nominal size and shape, and to random background
charges created by charged impurities.  Future work of interest may include
more precisely defined single-electron devices, using better fabrication
technology, and the extension of quantitative modeling to
semiconductor-based single-electronic circuits and hybrid single-electronic
/ conventional transistor logic circuits.  As single-electronics evolves
into a mature technology, such modeling will be essential.

\section{Acknowledgments}

We greatly appreciate numerous fruitful discussions with D. Averin,
R. Chen, L. Fonseca, A. Korotkov, W. Zheng, and K. Nabors.  This work was
supported in part by AFOSR grants \#F49620-1-0044 and \#F49620-96-1-0320.

\onecolumn
\begin{table}
\label{Simat}
\begin{tabular}{ddddddddddd}
 &6&7&8&9&10&11&12&13&14&15\\
\hline
1&1.8268&0.0304&0.0452&0.0176&0.0343&0.0139&0.0749&0.0569&0.0342&0.0417\\
2&0.0122&0.0060&0.0146&0.0073&0.0181&0.0092&0.0883&0.1418&0.0393&0.0630\\
3&0.0388&0.0168&0.0352&0.0150&0.0305&0.0125&0.0838&0.0564&2.0413&0.1169\\
4&0.0167&0.0074&0.0163&0.0072&0.0153&0.0065&0.0461&0.0360&2.0835&0.1529\\
5&0.0113&0.0053&0.0125&0.0059&0.0137&0.0064&0.0551&0.0599&0.0864&0.1889\\
\\
 6&3.7254\\
 7&-1.7587&3.6101\\
 8&-0.0223&-1.7587&3.7259\\
 9&-0.0038&-0.0084&-1.7587&3.6103\\
10&-0.0056&-0.0038&-0.0222&-1.7586&3.7261\\
11&-0.0016&-0.0012&-0.0037&-0.0084&-1.7585&3.6104\\
12&-0.0062&-0.0036&-0.0117&-0.0075&-0.0352&-1.7678&5.7693\\
13&-0.0032&-0.0019&-0.0052&-0.0033&-0.0096&-0.0118&-3.4910&3.9670\\
14&-0.0020&-0.0010&-0.0027&-0.0014&-0.0034&-0.0016&-0.0312&-0.0228&4.6482\\
15&-0.0018&-0.0009&-0.0024&-0.0012&-0.0031&-0.0015&-0.0277&-0.0349&-0.2780&0.9550\\
\end{tabular}
\caption{Capacitance matrix generated by ImageCap for circuit on Si
substrate.  Rows 1-5 belong to {\boldmath $\tilde C$}, rows 6-15 belong
to {\boldmath $C$}.  By convention, all $\tilde C_{ij} > 0$.  All
values in $10^{-16}$F.}
\end{table}

\begin{table}
\label{bimat}
\begin{tabular}{ddddddddddd}
 &6&7&8&9&10&11&12&13&14&15\\
\hline
1&1.2590&0.0179&0.0147&0.0092&0.0110&0.0072&0.0267&0.0245&0.0147&0.0196\\
2&0.0044&0.0034&0.0051&0.0039&0.0060&0.0049&0.0310&0.0580&0.0151&0.0254\\
3&0.0135&0.0095&0.0125&0.0086&0.0111&0.0074&0.0338&0.0270&1.2194&0.0540\\
4&0.0060&0.0043&0.0059&0.0041&0.0056&0.0038&0.0181&0.0166&1.2317&0.0638\\
5&0.0045&0.0033&0.0048&0.0035&0.0051&0.0037&0.0207&0.0248&0.0307&0.0739\\
\\
6&2.5498\\
7&-1.2370&2.5324\\
8&-0.0070&-1.2370&2.5501\\
9&-0.0017&-0.0067&-1.2370&2.5321\\
10&-0.0015&-0.0017&-0.0070&-1.2369&2.5500\\
11&-0.0006&-0.0009&-0.0017&-0.0066&-1.2369&2.5323\\
12&-0.0016&-0.0015&-0.0033&-0.0035&-0.0111&-1.2418&3.7851\\
13&-0.0009&-0.0010&-0.0016&-0.0019&-0.0034&-0.0084&-2.3386&2.5605\\
14&-0.0005&-0.0004&-0.0007&-0.0006&-0.0009&-0.0007&-0.0112&-0.0087&2.6644\\
15&-0.0005&-0.0004&-0.0007&-0.0006&-0.0009&-0.0007&-0.0100&-0.0145&-0.1097&0.4104\\
\end{tabular}
\caption{Capacitance matrix generated by ImageCap for circuit on SiO$_2$/Si
substrate.  Rows 1-5 belong to {\boldmath $\tilde C$}, rows 6-15 belong
to {\boldmath $C$}.  All values in $10^{-16}$F.}
\end{table}

\newpage
\begin{table}
\label{dV_Si}
\begin{tabular}{c|rrc|rrc}
&\multicolumn{3}{c}{Si substrate}&\multicolumn{3}{c}{SiO$_2$/Si substrate}\\
Node $i$~~&$\Delta V_i^{exp}$&$\Delta V_i^{sim}$&~~Difference~~~~&$\Delta V_i^{exp}$&$\Delta V_i^{sim}$&~~Difference\\
\hline
1&16(2)&8(1)&50\%&32(5)&20(2)&38\%\\
2&24(1)&17(2)&29\%&54(2)&45(5)&15\%\\
5&14(2)&7(1)&50\%&38(6)&51(5)&34\%\\
\end{tabular}
\caption{Values for $\Delta V_i$ in mV.}
\end{table}

\begin{table}
\begin{tabular}{l|ccc}
Substrate~~&$~\delta Q_o^{exp}$&$\delta Q_o^{sim}$&Difference\\
\hline
Si&0.064e&0.060e&6\%\\
SiO$_2$/Si&0.059e&0.045e&23\%\\
\end{tabular}
\caption{Response of electrometer to single electron entering the trap}
\end{table}

\end{document}